\documentclass[aps,prd,twocolumn,superscriptaddress,groupedaddress]{revtex4}

\usepackage{amssymb} \usepackage{color,graphicx} \usepackage{amsmath}
\usepackage{amsbsy} \usepackage{amsthm} 
\usepackage{bm} \usepackage{float} \usepackage{braket}
\usepackage{placeins}
\usepackage[colorlinks=true,citecolor=blue,linkcolor=red,urlcolor=red]{hyperref}
\usepackage[sort&compress]{natbib}
\usepackage{comment}

\newcommand{\bq}{\begin{equation}} \newcommand{\eq}{\end{equation}}
\newcommand{\bqali}{\begin{equation}\begin{aligned}} \newcommand{\eqali}{\end{aligned}\end{equation}}

\newcommand\D{\!\operatorname{d}\!}
\newcommand\x{{\bf x}}
\newcommand\vv{{\bf v}}
\newcommand\y{{\bf y}}
\newcommand\p{{\bf p}}
\newcommand\qq{{\bf q}}
\newcommand{\kB}{k_\text{\tiny B}}
\newcommand{\erf}{\,\operatorname{erf}\!}

\graphicspath{{../Figures/}}

\begin{document}

\author{Giovanni Di Bartolomeo}
\email{dibartolomeo.1419272@studenti.uniroma1.it}
\affiliation{Department of Physics, University of Rome "La Sapienza", Piazzale Aldo Moro 5, 00185 Rome, Italy}

\author{Matteo Carlesso}
\affiliation{Centre for Theoretical Atomic, Molecular, and Optical Physics, School of Mathematics and Physics, Queens University, Belfast BT7 1NN, United Kingdom}

\author{Angelo Bassi}
\affiliation{Department of Physics, University of Trieste, Strada Costiera 11, 34151 Trieste, Italy}
\affiliation{Istituto Nazionale di Fisica Nucleare, Trieste Section, Via Valerio 2, 34127 Trieste, Italy}

\title{Gravity as a classical channel and its dissipative generalization}

\date{\today}
\begin{abstract}
Recent models formulated by Kafri, Taylor, and Milburn and by Tilloy and Diosi describe the gravitational interaction through a continuous measurement and feedback protocol. In such a way, although gravity is ultimately treated as classical, they can reconstruct the proper quantum gravitational interaction at the level of the master equation for the statistical operator. Following this procedure, the price to pay is the presence of decoherence effects leading to an asymptotic energy divergence. One does not expect the latter in isolated systems. Here, we propose a dissipative generalization of these models. We show that, in these generalizations, in the long time limit, the system  thermalizes to an effective finite temperature.
\end{abstract}
 \maketitle

\section{Introduction}

The unification of the quantum theory with general relativity is still an open problem. The two theories well perform in their respective realms, yet one still needs to bridge an important gap in order to arrive at a unique theory.
While there are several approaches trying to quantize gravity \cite{borzeszkowski1988meaning,kiefer2007quantum,oriti2009approaches}, such as string theory \cite{green19871}, loop quantum gravity \cite{rovelli2000living} or spin foam quantum gravity \cite{steinhaus2020coarse}, a clear empirical evidence that gravity should be treated quantum mechanically is still lacking \cite{peres2001hybrid,wuthrich2005quantize,rothman2006can,carlip2008quantum,ahmadzadegan2016classicality,bose2017spin,marletto2017gravitationally,krisnanda2017revealing,belenchia2018icv,hall2018two,christodoulou2019possibility,carney2019tabletop,kumar2020quantum,donadi2021underground,marshman2020locality,bhole2020witnesses,chevalier2020witnessing,miao2020quantum,rijavec2021decoherence,rojas2021thermal,miki2021entanglement}. The other option is to move closer to the realm of general relativity as we know it, modifying quantum mechanics \cite{penrose1996gravity,diosi2014gravitation,diosi2011gravity,bowen2015quantum, bassi2017gravitational,khosla2017detecting}.
In the spirit of this latter approach, some proposed that gravity should be fundamentally classical and that the quantum dynamics should accommodate for it \cite{kafri2013noise, kafri2015bounds, tilloy2017principle, diosi2017gkls, altamirano2018gravity}. The price to pay in such a framework is the appearance of non-linear and stochastic terms in the Schr\"odinger equation, which lead to decoherence, and thus to the rupture of the energy conservation of an isolated system \cite{jacobs2006straightforward, wiseman2009quantum,kafri2014classical, tilloy2016sourcing, reyes2020gravitational}. While deviations from quantum mechanics are expected when moving toward the realm of general relativity, energy conservation is something one would want to maintain also in an hybrid model.

Here we analyze two models which develop a continuous measurement and feedback protocol to include classical Newtonian gravity in the quantum framework. Those models are the Kafri, Taylor and Milburn (KTM) model \cite{kafri2014classical} and the Tilloy-Diosi (TD) model \cite{tilloy2016sourcing}. Both feature a violation of energy conservation due to this protocol{, which is unexpected from the dynamics of isolated systems. 
}Here, we delve into the possibility of constructing a dissipative generalization of these two models. Under this perspective, the gravitationally-induced stochastic noise acts in an isolated system as a dissipative medium, similarly to what a thermal bath does in a typical open quantum system. In particular as we will see the energy of the system will reach an asymptotic finite value. 

The paper is structured as follows. In section \ref{due}, we briefly introduce the KTM model underling its violation of the energy conservation principle. In section \ref{tre}, we propose a dissipative extension of the KTM model, explicitly showing that it provides a finite asymptotic energy. In section \ref{quattro}, we review the TD model, which also does not conserve the total energy; then, we propose its dissipative extension. In section \ref{cinque}, we compare the dissipative TD model to the dissipative KTM model in the appropriate linear limit.

\section{The KTM model}\label{due}

To set the contest of the problem, we briefly introduce the KTM model highlighting the relevant features.
The model consists of a one-dimensional system composed of two masses $m_{1}$ and $m_{2}$, which
 are harmonically trapped at frequencies $\omega_1$ and $\omega_2$ at a distance $d$, and interact gravitationaly. Assuming that the quantum fluctuations in position $|\hat x_1-\hat x_2|$, with $\hat x_i$ the position operator of the $i$-th particle, are small compared to $d$, one can approximate the Newtonian potential to the second order in $\hat x_1-\hat x_2$. 
With a suitable choice of the coordinates \cite{kafri2014classical}, the Hamiltonian reads $\hat H =  \hat H_0 + \hat{H}_{\text{grav}}$,
where $\hat{H}_{0} = \sum_{k=1}^{2}(\frac{\hat{p}_{k}^{2}}{2m_{k}} + \frac 12 m_{k}\Omega_{k}^{2}\hat{x}_{k}^{2})$ and
\begin{equation}
\hat{H}_{\text{grav}} = K\hat{x}_{1}\hat{x}_{2},
\end{equation}
with $\Omega_{k}^{2} = \omega_{k}^{2} - {K}/{m_{k}}$ and we have defined $K = {2Gm_{1}m_{2}}/{d^{3}}$ where $G$ is the gravitational constant.

The key idea of the model is that the \textit{quantum} interaction $\hat H_\text{grav}$ between the two masses is replaced by a \textit{classical} protocol. The latter consists of the continuous weak measurement of the positions $\hat{x}_{k}$ of each particle and the broadcast of the corresponding measurement record $r_k$ to the other particle through a classical channel. The gravitational interaction is realized through a feedback dynamics, which is implemented by replacing $\hat{H}_{\text{grav}}$ with the feedback Hamiltonian
\begin{equation}\label{HfbKTM}
\hat{H}_{\text{fb}} = \chi_{1}r_{1}\hat{x}_{2} + \chi_{2}r_{2}\hat{x}_{1},    
\end{equation}
where the position operator of one mass is coupled to the classical stochastic measurement record of the position of the other mass. In particular, the measurement record is defined as
\begin{equation}
r_{k} = \braket{\hat{x}_{k}}_{t} + \frac{\hbar}{\sqrt{\gamma_{k}}}\frac{\D W_{k,t}}{\D t},
\end{equation}
where $\gamma_{k}$ are the measurement information gain rates and $W_{k,t}$ are the standard Wiener processes, whose correlations read $\mathbb E[\,\D W_{k,t}\D W_{l,t}]=\delta_{k,l}\D t$. Following the calculations reported in detail in Appendix~\ref{appendixA}, one arrives at the following non-linear and stochastic equation for the state vector $\ket{\psi_{t}}$
\bqali
\label{continuousmeasure}
& \D\ket{\psi_{t}} = \Biggl\{-\sum_{{\substack{k,j=1\\j\ne k}}}^{2} \frac{i}{2\hbar}\chi_{k} \hat{x}_{j} (\hat{x}_{k} - \braket{\hat{x}_{k}}_{t})\D t \\
&+\sum_{k=1}^{2}\biggl[-\frac{\gamma_{k}}{8\hbar^{2}}\left(\hat{x}_{k} - \braket{\hat{x}_{k}}_{t}\right)^{2}\D t +\frac{\sqrt{\gamma_{k}}}{2\hbar}\left(\hat{x}_{k} -\braket{\hat{x}_{k}}_{t}\right)\D W_{k,t}\biggr]\\
&-\!\!\sum_{{\substack{k,j=1\\j\ne k}}}^{2}\!\! \left[\left( \frac{i\chi_{k}\braket{\hat{x}_{k}}_{t} \hat{x}_{j}}{\hbar}  + \frac{\chi_{k}^{2}\hat{x}_{j}^{2}}{2\gamma_{k}}  \right) \D t \right.\left. -\frac{i\chi_{k}\hat{x}_{j}}{\sqrt{\gamma_{k}}} \D W_{k,t}\right]\Biggr\}\ket{\psi_{t}}.
\eqali
Here, the second line is given by the continuous measurement process, the last line is the feedback contribution, while the first one arises from the combined effect of the two processes. We stress again that at this level gravity enters in a semi-classical, non-linear and stochastic manner, with no apparent resemblance with the (linearized) Newtonian potential usually entering the Schr\"odinger equation.
By setting $\chi_{1} = \chi_{2} = K$, we find the corresponding KTM master equation \cite{kafri2014classical}:
\begin{equation}
\label{kafrinotminimum}
    \frac{\D\hat\rho_{t}}{\D t} = -\frac{i}{\hbar}[\hat{H}_{0} + K\hat{x}_{1}\hat{x}_{2},\hat\rho_{t}] -\!\! \sum_{{\substack{k,j=1\\j\ne k}}}^{2}\!\! \left(\frac{\gamma_{k}}{8\hbar^{2}} + \frac{K^{2}}{2\gamma_{j}}\right) [\hat{x}_{k},[\hat{x}_{k},\hat\rho_{t}]],
\end{equation}
where $\hat\rho_{t} = \mathbb E[\ket{\psi_{t}}\bra{\psi_{t}}]$ and we added the free evolution described by $\hat{H}_{0}$.
Equation~\eqref{kafrinotminimum} comprises two terms: the first is a von Neumann term, where the usual gravitational interaction of non-relativistic quantum mechanics is reproduced, while the second is a decoherence term which originates from the stochastic dynamics induced by the continuous measurement and feedback mechanism.
Now, if we set $m_{k} = m$, it is then reasonable to consider $\gamma_{k}  = \gamma$. The parameter $\gamma$ is free, but it can be suitably fixed to minimize the corresponding decoherence effects. After such a minimization, corresponding to $\gamma=\gamma_\text{\tiny KTM}$ with 
\bq\label{gammaKTM}
\gamma_\text{\tiny KTM}=2\hbar K,
\eq
we obtain the following master equation \cite{kafri2014classical}:
\begin{equation}
\label{kafri}
\frac{\D}{\D t}\hat\rho_{t} = -\frac{i}{\hbar}[\hat{H}_{0} + K\hat{x}_{1}\hat{x}_{2},\hat\rho_{t}] - \frac{K}{2\hbar} \sum_{k=1}^{2} [\hat{x}_{k},[\hat{x}_{k},\hat\rho_{t}]].
\end{equation}
The second term quantifies the minimum decoherence effect induced by the protocol, which is not zero.
Among the predictions of  Eq.~\eqref{kafri}, and Eq.~\eqref{kafrinotminimum} as well, one has that the mean energy of the system increases linearly in time, eventually diverging in the long-time limit. {Indeed, independently from the details of the potential in $\hat H_0$, the contribution of the second term of Eq.~\eqref{kafri} to the single-particle kinetic energy $\tfrac{1}{2m}\braket{\hat p_k^2}$ gives $\operatorname{Tr}\{-\tfrac{K}{2\hbar}[\hat x,[\hat x,\frac{1}{2m}\hat p^2]\hat \rho\}=K \hbar/2m$. 
%
Consequently}, for a system of two identical { harmonically-trapped} masses $m$, the expectation value of the Hamiltonian $\hat H$ reads 
\begin{equation}
\braket{\hat{H}}_{t} = \frac{\hbar K}{m}t,
\end{equation}
which grows linearly in time.

In this work we show how it is possible to modify the measurement and feedback protocol keeping the energy bounded, while reproducing the correct quantum gravitational interaction in the von Neumann term of the master equation.

\section{The Dissipative KTM model}\label{tre}
To avoid the asymptotic divergence of the average energy arising in the KTM model, we propose a dissipative generalization in analogy to the quantum Brownian model \cite{caldeira1983path,Hu:1992va, Ford:2001uw,breuer2002theory,gardiner2004quantum,schlosshauer2007decoherence,weiss2012quantum, joos2013decoherence,Carlesso:2017ue}. The latter describes the motion of a massive harmonic oscillator under the influence of a thermal environment. When extending the quantum Brownian model to two particles having mass $m_{k}$ and frequency $\omega_{k}$, the master equation reads
\cite{caldeira1983path, vacchini2000completely} 
\bqali
\label{caldeira}
&\frac{\D}{\D t}\hat{\rho}  = -\frac{i}{\hbar}[\hat{H},\hat{\rho}] - \sum_{k=1}^{2} \frac{i\lambda_{k}}{\hbar} [\hat{x}_{k},\{\hat{p}_{k},\hat{\rho}\}] \\
& - \sum_{k=1}^{2} \frac{2\lambda_{k}m_{k}\kB T}{\hbar^{2}}[\hat{x}_{k},[\hat{x}_{k},\hat{\rho}]] - \sum_{k=1}^{2} \frac{\lambda_{k}}{8m_{k}\kB T}[\hat{p}_{k},[\hat{p}_{k},\hat{\rho}]],
\eqali
where $\lambda_{k}$ are the dissipative constants and $T$ is the temperature of the bath, { and we used the standard notation for anticommutator $\{\hat{A},\hat{B}\} = \hat{A}\hat{B} + \hat{B}\hat{A}$}. In the high temperature limit the last term of Eq.~\eqref{caldeira} becomes negligible, and the asymptotic average energy of the system reads 
\begin{equation}
\label{Hs}
\braket{\hat{H}}_{\infty} = 2\kB T.
\end{equation}
The system thermalizes to a finite energy, which is in agreement with the equipartition theorem in the canonical statistical ensemble \cite{huang2009introduction}. We underline that $T$ in the canonical ensemble is a universal temperature, namely it does not depend on the specific properties of the system but only on those of the bath. Thus, when generalizing the KTM model to include dissipative features, it is desirable to have an asymptotic energy which is independent from the specific properties of the system. Conversely, such asymptotic value should be analogue to the temperature in the quantum Brownian motion.

We include dissipative effects in the KTM model by modifying the measurement and feedback protocol so that the corresponding master equation is similar to Eq.~\eqref{caldeira}. Clearly, in this modificaton, we have to preserve the main properties of the original model, namely we need to recover the linearized gravitational interaction. This can be done by substituting the continuous measurement of $\hat x_k$ with that of the following operator
\begin{equation}
\label{ourchoice}
\hat{A}_{k} = \hat{x}_{k} + \frac{i\alpha_{k}}{\hbar}\hat{p}_{k}, 
\end{equation}
where $\alpha_{k}$ are real parameters to be determined. 
Conversely, we do not modify the form of the feedback Hamiltonian, which will continue to read as in Eq.~\eqref{HfbKTM}, where now the measurement record reads
\bq\label{m.record}
r_{k} = \frac{1}{2}\braket{\hat{A}_{k} + \hat{A}_{k}^{\dagger}}_{t} + \frac{\hbar}{\sqrt{\gamma_{k}}}\frac{\D W_{k,t}}{\D t}.
\eq
This choice is twofold: we obtain the same dissipative term as that in Eq.~\eqref{caldeira}, which depends on both the position and momentum operators, and we mimick the gravitational Hamiltonian $\hat{H}_\text{grav}$. 
{ The latter result is determined by the fact that $\tfrac12(\hat A_k+\hat A_k^\dag)=\hat x_k$ -- consequently the feedback Hamiltonian and the measurement record are the same as in the KTM model -- and thus the gravitational interaction is correctly reproduced at the linear order. The former result  instead is possible due to the different  choice of the measured operator in  Eq.~\eqref{ourchoice}, which drives the continuous measurement and allows to introduce the desired dissipative effects.}
Now, by taking
$\chi_{1} = \chi_{2} = K$, we straightforwardly arrive at the following master equation [cf.~Appendix~\ref{appendixA}]:
\bqali
\label{dissipationrho}
&\frac{\D}{\D t}\hat\rho_{t}  = -\frac{i}{\hbar}[\hat{H}'_{0} + K\hat{x}_{1}\hat{x}_{2} ,\hat\rho_{t}] - \sum_{k=1}^{2} \frac{i\gamma_{k}\alpha_{k}}{4\hbar^{3}} [\hat{x}_{k},\{\hat{p}_{k},\hat\rho_{t}\}]\\
& - \sum_{{\substack{k,j=1\\j\ne k}}}^{2} \left(\frac{\gamma_{k}}{8\hbar^{2}} + \frac{K^{2}}{2\gamma_{j}}\right) [\hat{x}_{k},[\hat{x}_{k},\hat\rho_{t}]]  - \sum_{k=1}^{2} \frac{\gamma_{k}\alpha_{k}^{2}}{8\hbar^{4}}[\hat{p}_{k},[\hat{p}_{k},\hat\rho_{t}]] \\
& + \sum_{{\substack{k,j=1\\j\ne k}}}^{2} \frac{\alpha_{j}K}{2\hbar^{2}} [\hat{x}_{k},[\hat{p}_{j},\hat\rho_{t}]],
\eqali
where $\hat{H}'_{0} = \hat{H}_{0} + \Delta\hat{H}_{0}$, with $\Delta\hat{H}_{0} = -\sum_{k=1}^{2}\frac{\gamma_{k}\alpha_{k}}{8\hbar^{2}}\{\hat{x}_{k},\hat{p}_{k}\}$ being an addition to the Hamiltonian deriving from the continuous measurement.

We now compare Eq.~\eqref{dissipationrho} and Eq.~\eqref{caldeira}.
First, we notice that the effective Hamiltonian in the first term of Eq.~\eqref{dissipationrho} comprises the quantum gravitational interaction as in the KTM master equation \eqref{kafri}. Moreover,  Eq.~\eqref{dissipationrho} displays three terms analogous to those in Eq.~\eqref{caldeira}, and which implement dissipation, diffusion in momentum and in position respectively.
Conversely to Eq.~\eqref{caldeira}, Eq.~\eqref{dissipationrho} contains also a new term, the last, which derives from  the feedback mechanism and also produces diffusion in position and momentum. 

\subsection{Asymptotic energy of the KTM model}

To verify that our modification can actually solve the energy divergence problem, we start by computing 
the asymptotic value of the average energy in the simple case of $\gamma_{k} = \gamma_\text{\tiny KTM}$, $m_{k} = m$ and $\alpha_{k} = \alpha$. For the sake of simplicity, we will assume $\hat{H}'_{0} = \hat{H}_{0}$ in Eq.~\eqref{dissipationrho}, since no substantial change in the mechanism causing the thermalization of the system is expected. To simplify the calculations, we move to center-of-mass and relative displacement coordinates, which are defined as
$\hat{x}_\text{cm}  = \frac{1}{2}(\hat{x}_{1} + \hat{x}_{2})$, $\hat{p}_\text{cm} = \hat{p}_{1} + \hat{p}_{2}$, $\hat{x}_\text{rel} = \hat{x}_{1} - \hat{x}_{2}$ and $ \hat{p}_\text{rel} = \frac{1}{2}(\hat{p}_{1} - \hat{p}_{2})$.
Then, the Hamiltonian in Eq.~\eqref{dissipationrho} can be rewritten as $
\hat{H} = \hat{H}_\text{cm} + \hat{H}_\text{rel}$,
where
\bqali
\hat{H}_\text{cm} = \frac{\hat{p}_\text{cm}^{2}}{4m} + m\omega^{2}\hat{x}_\text{cm}^{2}\quad \text{and} \quad
\hat{H}_\text{rel} = \frac{\hat{p}_\text{rel}^{2}}{m} + \frac{m}{4}\tilde\Omega^{2} \hat{x}_\text{rel}^{2},
\eqali
are respectively the Hamiltonian center-of-mass of mass $2m$, and that of the relative displacement with mass $m/2$ and frequency $\tilde\Omega=\sqrt{\Omega^2-K/m}$. Consequently, by assuming that $\gamma_k$ takes the expression in Eq.~\eqref{gammaKTM}, we find that Eq.~\eqref{dissipationrho} can be divided in two independent master equations. The one for the center-of-mass reads
\bqali
\label{centerofmassKTM}
&\frac{\D}{\D t}\hat{\rho}_\text{cm}  = -\frac{i}{\hbar}[\hat{H}_\text{cm},\hat{\rho}_\text{cm}] - \frac{K}{\hbar}[\hat{x}_\text{cm},[\hat{x}_\text{cm},\hat{\rho}_\text{cm}]] \\
& - \frac{iK\alpha}{2\hbar^{2}} [\hat{x}_\text{cm},\{\hat{p}_\text{cm},\hat{\rho}_\text{cm}\}] - \frac{K\alpha^{2}}{8\hbar^{3}}[\hat{p}_\text{cm},[\hat{p}_\text{cm},\hat{\rho}_\text{cm}]] \\
& + \frac{\alpha K}{2\hbar^{2}} [\hat{x}_\text{cm},[\hat{p}_\text{cm},\hat{\rho}_\text{cm}]],
\eqali
while that for  the relative displacement is given by
\bqali
\label{relativeKTM}
&\frac{\D}{\D t}\hat{\rho}_\text{rel}  = -\frac{i}{\hbar}[\hat{H}_{\text{rel}},\hat{\rho}_{\text{rel}}] - \frac{K}{4\hbar}[\hat{x}_{\text{rel}},[\hat{x}_{\text{rel}},\hat{\rho}_{\text{rel}}]] \\
& - \frac{iK\alpha}{2\hbar^{2}} [\hat{x}_{\text{rel}},\{\hat{p}_{\text{rel}},\hat{\rho}_{\text{rel}}\}] - \frac{K\alpha^{2}}{2\hbar^{3}}[\hat{p}_{\text{rel}},[\hat{p}_{\text{rel}},\hat{\rho}_\text{rel}]] \\
& - \frac{\alpha K}{2\hbar^{2}} [\hat{x}_{\text{rel}},[\hat{p}_{\text{rel}},\hat{\rho}_{\text{rel}}]].
\eqali
By defining $\hat{T}_{cm} = {\hat{p}_\text{cm}^{2}}/{4m}$, $\hat{V}_\text{cm} = m\omega^{2}\hat{x}_{cm}^{2}$, $\hat{T}_{\text{rel}} = {\hat{p}_{\text{rel}}^{2}}/{m}$ and $\hat{V}_{\text{rel}} = {m}\tilde\Omega^{2} \hat{x}_{\text{rel}}^{2}/4$,  we obtain, through Eq.~\eqref{centerofmassKTM} and Eq.~\eqref{relativeKTM},  two systems of three coupled differential equations of the first order. The system for the center-of-mass reads
\bqali
\label{centermassdiff}
&\frac{\D}{\D t}\braket{\hat{V}_\text{cm}}_{t}  = \frac{\omega^{2}}{2}\braket{\{\hat{p}_\text{cm},\hat{x}_\text{cm}\}}_{t} + \frac{K\alpha^{2}m\omega^{2}}{4\hbar}, \\
&\frac{\D}{\D t}\braket{\hat{T}_\text{cm}}_{t}  = - \frac{2K\alpha}{\hbar}\braket{\hat{T}_\text{cm}}_{t} - \frac{\omega^{2}}{2}\braket{\{\hat{p}_\text{cm},\hat{x}_\text{cm}\}}_{t}  + \frac{\hbar K}{2m}, \\
&\frac{\D}{\D t}\braket{\{\hat{p}_\text{cm},\hat{x}_\text{cm}\}}_{t} = - \frac{K\alpha}{\hbar}\braket{\{\hat{p}_\text{cm},\hat{x}_\text{cm}\}}_{t} + 4\braket{\hat{T}_\text{cm}}_{t} \\
& - 4\braket{\hat{V}_\text{cm}}_{t} + \alpha K,
\eqali
while that for the relative degrees of freedom is
\bqali
\label{relativediff}
&\frac{\D}{\D t}\braket{\hat{V}_{\text{rel}}}_{t}  = \frac{1}{2}\tilde\Omega^{2} \braket{\{\hat{p}_{12},\hat{x}_{\text{rel}}\}}_{t} 
+ \frac{K\alpha^{2}m}{4\hbar}\tilde\Omega^{2},  \\
&\frac{\D}{\D t}\braket{\hat{T}_{\text{rel}}}_{t}  = - \frac{2K\alpha}{\hbar}\braket{\hat{T}_{\text{rel}}}_{t} - \frac{1}{2}\tilde\Omega^{2}\braket{\{\hat{p}_{\text{rel}},\hat{x}_\text{rel}\}}_{t} + \frac{\hbar K}{2m}, \\
&\frac{\D}{\D t}\braket{\{\hat{p}_{12},\hat{x}_{\text{rel}}\}}_{t}  = - {K\alpha}(1+\frac{\braket{\{\hat{p}_{\text{rel}},\hat{x}_{\text{rel}}\}}_{t}}{\hbar}) + 4\braket{\hat{T}_{\text{rel}}-\hat{V}_{\text{rel}}}_{t}.
\eqali
Finally, we can obtain the asymptotic energy of the system by setting the above derivatives to zero. Thus, once summing the four contribution to $\hat H=\hat{T}_\text{cm} + \hat{V}_\text{cm}+\hat{T}_{\text{rel}} + \hat{V}_{\text{rel}}$ we find
\begin{equation}
\label{asintototappendix}
\braket{\hat{H}}_{\infty} = \frac{\hbar^{2}}{m\alpha} + \frac{\alpha m \Omega^{2}}{2} + \frac{K^{2}\alpha^{3}m}{4\hbar^{2}},
\end{equation}
By  making explicit all the constants and parameters of the model we have
\begin{equation}
\label{energiaasint}
\braket{\hat{H}}_{\infty} = \frac{\hbar^{2}}{m\alpha} + \frac{\alpha m \omega^{2}}{2} -\frac{\alpha m^{2}G}{d^{3}} + \frac{G^{2}\alpha^{3}m^{5}}{\hbar^{2}d^{6}}.
\end{equation}
We notice that Eq.~\eqref{energiaasint} depends on the parameters of the system, namely the mass $m$, the distance $d$ and the frequency $\omega$, and on the free parameter $\alpha$ of the model. On the contrary, in order to associate a universal temperature to the system, one would expect an expression free of such dependencies, as that in Eq.~\eqref{Hs}.
A way to approximately remove such dependence is to assume that $\alpha$ is suitably small in such a way to retain only the first term of Eq.~\eqref{energiaasint}. Then, by defining
\begin{equation}
\label{alfa}
\alpha = \frac{m_{0}}{m}\alpha_{0},
\end{equation}
where $m_{0}$ is a reference mass 
and $\alpha_{0}$ is a free parameter, the asymptotic average energy becomes
\begin{equation}
\label{energyapproxasympt}
\braket{\hat{H}}_{\infty} = \frac{\hbar^{2}}{m_{0}\alpha_{0}},
\end{equation}
which is system independent.
By comparing such an expression with that in Eq.~\eqref{Hs}, we can define a temperature at which the system will eventually thermalizes. This reads
\begin{equation}
\label{tefft}
T_{\text{eff}} = \frac{\hbar^{2}}{2m_{0}\alpha_{0}\kB}.
\end{equation}
Thus, in light of the analogy with the quantum Brownian model, we can interpret the dynamics described by  the dissipative KTM model as that of a system in contact with a thermal bath of temperature $T_{\text{eff}}$. Here the bath is associated to the measurement process and feedback protocol.

We now inquire what changes without making the choice $\gamma = \gamma_\text{\tiny KTM}$.
In this case the expression in Eq.~\eqref{energiaasint} without such an assumption, becomes
\begin{equation}
\label{Tnotminimized}
\braket{\hat{H}}_{\infty}  = \frac{\hbar^{2}}{2m\alpha} +\frac{8\hbar^{4}G^2m^{3}}{\gamma^{2}\alpha d^{6}}+\frac{\alpha m \omega^{2}}{2} - \frac{\alpha m^{2}G}{d^{3}}+\frac{m\alpha^{3}\gamma^{2}}{16\hbar^{4}}.
\end{equation}
By choosing $\alpha = \frac{m_{0}}{m}\alpha_{0}$, we remove the dependence on the mass in the first and third term of Eq. \eqref{Tnotminimized}. Thus, in order to remove the dependence on the mass and on the distance of the second term, we can choose for example $\gamma =\frac{m^{2}}{m_{0}^{2}d^{3}}\gamma_{0}$. However, in this way, one still has the dependence on the frequency in the third term and that on the mass and on the distance in the fourth and fifth term.
If the values of $\alpha$ and $\gamma$ are both suitably small we can neglect the last three terms in Eq.~\eqref{Tnotminimized} and find
\begin{equation}
\label{temperaturacongamma}
\braket{\hat{H}}_{\infty} = \frac{\hbar^{2}}{2m_{0}\alpha_{0}} +\frac{4\hbar^{4}G^2m_{0}^{3}}{\gamma_{0}^{2}\alpha_{0}}.
\end{equation}
In such a way the effective temperature reads
\begin{equation}
\label{temperaturacongammaapprox}
T_{\text{eff}} = \frac{\hbar^{2}}{4m_{0}\alpha_{0}\kB} +\frac{2\hbar^{2}G^2m_{0}^{3}}{\gamma_{0}^{2}\alpha_{0}\kB}.
\end{equation}
We notice that Eq.~\eqref{temperaturacongammaapprox} depends explicitly on the gravitational constant $G$ unlike Eq.~\eqref{tefft}. The first term of Eq.~\eqref{temperaturacongammaapprox} is due to the measurement process, while the second term is due to the feedback mechanism. Thus, by comparing Eq.~\eqref{temperaturacongammaapprox} with Eq.~\eqref{tefft} is clear that the effect of the choice $\gamma = \gamma_\text{\tiny KTM}$ is to make the contributions of the measurement and of the feedback indistinguishable.

\section{The TD model and its dissipative generalization}\label{quattro}
The second model we consider is the Tilloy-Diosi (TD) model \cite{tilloy2016sourcing}. Similarly to the KTM model, also here a weak continuous measurement is performed with the subsequent classical broadcast of the corresponding measurement record, which modifies the system dynamics through a feedback Hamiltonian. The conceptual difference lies in the way the Newtonian gravitational interaction is implemented \cite{reyes2020gravitational}. While in the KTM model the gravitational interaction is approximated to the linear regime, the full Newtonian potential is
\begin{equation}\label{eq.newton}
\hat{H}_\text{grav} = \frac{1}{2}\int \D^{3}x\int \D^{3}y\, V(\x - \y)\hat\mu(\x)\hat{\mu}(\y),
\end{equation}
where $V({\x} - {\y}) = -G / |{\x} - {\y}|$ is the Newtonian potential.
In particular, what is measured here is the mass density $\hat{\mu}(\x)$ and the corresponding feedback Hamiltonian reads
\begin{equation}
\hat{H}_{\text{{fb}}} = \int \D^{3}x\int\D^{3}y\,V({\x} - {\y}) \hat{\mu}({\x})r({\y}),    
\end{equation}
where
\begin{equation}
\label{recordTD}
r({\x}) = \braket{\hat{\mu}({\x})}_{t} + \hbar\int\D^{3}y\,\gamma^{-1}({\x} - {\y})\frac{\D W_{t}({\y})}{\D t},
\end{equation}
is the measurement record of the mass density, where $\gamma(
{\x} - \y)$ is a spatial correlation function, $\gamma^{-1}(
{\x} - \y)$ its inverse function and $W_{t}(\x)$ is a standard Wiener process with zero average and correlations $\mathbb{E}[\D W_{t}(\x)\D W_{t}(\y)] = \gamma(\x - \y)\D t$. By following the procedure highlighted in Appendix \ref{appendixA}, one can derive the non-linear and stochastic equation for the state vector $\ket{\psi_t}$ of the system, which will take a form analogous to that in Eq.~\eqref{continuousmeasure}. Then, one derives the corresponding master equation, which reads
\cite{tilloy2016sourcing}
\bqali
\label{TDrho1}
\frac{\D}{\D t}\hat\rho_{t} &= -\frac{i}{\hbar}[\hat{H}_{0} + \hat{H}_\text{grav},\hat\rho_{t}] \\
& + \frac{1}{2\hbar}\int \D^{3}x\int\D^{3}y\,V(\x - \y) [\hat{\mu}(\x),[\hat{\mu}(\y), \hat\rho_{t}]],
\eqali
where we chose $\gamma(\x - \y) = - 2\hbar V(\x - \y)$ and $\hat H_0$ is the free Hamiltonian. As in KTM model, also in TD model the gravitational interaction $\hat{H}_\text{grav}$ is reproduced in the von Neumann term, although one pays the price of having an additional gravitational decoherence term. 

We notice that due to the form of the Newtonian gravitational potential, the integrals in Eq.~\eqref{TDrho1} are in general divergent. However, they can be regularized by using a suitable smearing function $g(\x)$. Here, we will consider a normalized Gaussian smearing of the form
\begin{equation}
\label{gaussiansmearing}
g(\x - \y) = \frac{e^{-\frac{|\x - \y|^{2}}{2R_{0}^{2}}}}{(2\pi R_{0}^{2})^{3/2}},
\end{equation}
where $R_0$ sets its variance. The latter can be interpret as the minimum gravitational interaction distance~\cite{diosi2014gravitation, penrose1996gravity}, and thus becomes an extra  parameter of the model.

As for the KTM model, also for the TD model the asymptotic energy in general is divergent. For example, if we consider a system of $N$ point-like particles, whose mass, position operator and mass density respectively read $m_k$, $\hat \x_k$ and 
\begin{equation}
\label{massdR}
\hat{\mu}(\x) = \sum_{k = 1}^{N}m_{k}\,\delta(\x - \hat{\x}_{k}),
\end{equation}
we find that Eq.~\eqref{TDrho1} gives
\bq
\braket{\hat H}_t=\frac{\hbar G \sum_k m_k}{4\sqrt{\pi}R_{0}^{3}}t,
\eq
which, again, grows linearly in time.

To solve this issue, we aim at modifying the TD model by adding dissipative terms although still reproducing the quantum gravitational interaction. This can be done through a specific choice of the operators to be measured. 
Two options are possible. The first one is to choose a suitable smearing of the mass density $\hat\mu(\x)$ to include a momentum operator. This is the approach that was used to construct the dissipative generalization of the Continuous Spontaneous Localization model \cite{smirne2015dissipative}. 
Following these lines, we consider
\bq
\hat A(\x) \!=\!\! \sum_{k=1}^{N}\!\frac{m_{k}}{(2\pi\hbar)^{3}}\!\!\int \D^{3}q\, e^{-\frac{i}{\hbar}\qq\cdot(\x - \hat\x_{k})-\frac{R_{0}^{2}}{2\hbar^{2}}[(1+\alpha_{k})\qq+2\alpha_{k}\hat{\p}_{k}]^{2}},
\eq
where $\alpha_{k}$ are real parameters.
With this measurement operator, we find that the measurement process provides the expected dissipative terms.
{However, with such a choice one cannot reconstruct the potential in Eq.~\eqref{eq.newton}; the resulting potential is:
\begin{equation}
\hat{H}_\text{I} =\frac12 \int {\D^{3}x\D^{3}y} \,V(\x - \y)\left(\hat{\mu}(\x)\hat{A}(\y) + \hat{A}^{\dagger}(\y)\hat{\mu}(\x)\right)
\end{equation}
which 
contains the momentum operators as well. Correspondingly, the equations of motion for $\hat \x_k$ and $\hat \p_k$ change, making the dynamics different from the one described by the Newtonian gravitational potential. To avoid this, we need to consider another  form of the measurement operator.
}

The second choice is to modify the measured operator by adding to the density operator a non-Hermitian part, similarly as we did with the position operator when constructing the dissipative KTM model in Sec.~\ref{tre}.
Following this idea, we consider the following operator
\begin{equation}
\label{ourchoiceTD}
\hat{A}(\x) = \hat{\mu}(\x) + i\hat{\mu}_\text{I}(\x),
\end{equation}
in place of the mass density of the system $\hat{\mu}(\x)$ alone. Here, $\hat{\mu}_\text{I}(\x)$ is an arbitrary Hermitian operator yet to be determined. 
The corresponding measurement 
can be computed through 
\begin{equation}
\label{recordsistemacontinuo}
r(\x) = \frac{1}{2}\braket{\hat{A}(\x) + \hat{A}^{\dagger}(\x)}_{t} + \hbar\int\D^{3}y\,\gamma^{-1}(\x - \y)\frac{\D W_{t}(\y)}{\D t},
\end{equation}
and is equal to that in Eq.~\eqref{recordTD}. Correspondingly, also the feedback Hamiltonian does not change with respect to that of the TD model, and thus the quantum gravitational interaction is correctly reproduced. Finally, by following the calculations reported in Appendix \ref{appendixA}, one derives the master equation of the dissipative generalization of the TD model, which reads
\bqali
\label{TDmasterdissip1}
\frac{\D}{\D t}\hat\rho_{t} &= - \frac{i}{\hbar}[\hat{H}'_{0} + \hat{H}_\text{grav},\hat\rho_{t}] \\
& + \frac{1}{2\hbar}\int \D^{3}x\int \D^{3}y\,V(\x - \y)[\hat{\mu}(\x),[\hat{\mu}(\y),\hat\rho_{t}]] \\
& +\frac{i}{2\hbar}\int \D^{3}x\int \D^{3}y\, V(\x - \y)[\hat{\mu}(\x),\{\hat{\mu}_\text{I}(\y),\hat\rho_{t}\}] \\
& +\frac{1}{4\hbar}\int \D^{3}x\int\D^{3}y\, V(\x - \y)[\hat{\mu}_\text{I}(\x),[\hat{\mu}_\text{I}(\y),\hat\rho_{t}]] \\
& +\frac{1}{2\hbar}\int \D^{3}x\int\D^{3}y\,V(\x - \y)[\hat{\mu}(\x),[\hat{\mu}_\text{I}(\y),\hat\rho_{t}]],
\eqali
where we set $\gamma(\x - \y) = - 2\hbar V(\x - \y)$  and we defined 
$\hat{H}'_{0} = \hat{H}_{0} + \Delta\hat{H}_{0}$,
where 
\begin{equation}
\label{further}
\Delta\hat{H}_{0} = -\frac{1}{4\hbar}\int \D^{3}x\int\D^{3}y\, V(\x - \y)\{\hat{\mu}(\x),\hat{\mu}_\text{I}(\y)\},
\end{equation}
is an additional term due to the continuous measurement process.
We stress that the structure of Eq.~\eqref{TDmasterdissip1} is analogous to that of the dissipative KTM model in Eq.~\eqref{dissipationrho}. 
In the following, for the sake of simplicity, we approximate $\hat{H}'_{0}$ to $\hat{H}_{0}$, since we expect that this does not substantially change the mechanism that causes the asymptotic mean energy of the system to be finite.

\section{Linear limit of the TD model}\label{cinque}

At this point of the discussion, we need to fix the form of  $\hat{\mu}_\text{I}(\x)$ to properly derive the desired dissipative dynamics. We start by considering a $N$ point-particle system, whose mass density is given in Eq.~\eqref{massdR}. We rewrite the position and momentum operators as
\begin{equation}
\label{fluttuazioni}
\hat{\x}_{k} = \x_{k}^{(0)} + \Delta\hat{\x}_{k} \quad \text{and} \quad \hat{\p}_{k} = \p_{k}^{(0)} + \Delta\hat{\p}_{k},
\end{equation}
where $\Delta\hat{\x}_{k}$ and $\Delta\hat{\p}_{k}$ are the quantum fluctuations with respect to the classical position $\x_{k}^{(0)}$ and  momentum $\p_{k}^{(0)}$ respectively. Then, we choose the following form for $\hat \mu_\text{I}(\x)$:
\begin{equation}
\label{massdI}
\hat{\mu}_\text{I}(\x) = \sum_{k = 1}^{N}m_{k}\, \delta\left(\x - \x_{k}^{(0)}-\frac{\alpha_{k}}{\hbar}\Delta\hat{\p}_{k}\right),
\end{equation}
where $\alpha_{k}$ are real free parameters yet to be determined. {The drive for the choice  in Eq.~\eqref{massdI} is that the resulting  dynamics satisfies  translational invariance. Indeed, it is straightforward to check that other choices for $\hat \mu_\text{I}(\x)$ of the form $\sum_k m_k\delta\left(\x - \vv_k-\frac{\alpha_{k}}{\hbar}\Delta\hat{\p}_{k}\right)$ would lead to the violation of the translational invariance for any choice of $\vv_k$ different from $\x_k^{(0)}$.
On the other hand, the dynamics is not boost invariant. This is however a common feature of the dissipative models, such as the quantum Brownian motion \cite{breuer2002theory}, the dissipative Continuous Spontaneous Localization model \cite{smirne2015dissipative} and the dissipative KTM model introduced in section \ref{tre}}.

Being point-like, these choices for $\hat \mu(\x)$ and $\hat{\mu}_\text{I}(\x)$ lead to  divergences in the master equation \eqref{TDmasterdissip1}, which are expected, similarly as those present in the TD model (see also the discussion after~Eq.~\eqref{TDrho1} and in Ref.~\cite{reyes2020gravitational}). We proceed then with a regularization of the gravitational potential $V(\x-\y)$, namely we implement the following substitution: $V(\x - \y) \to (g\ \circ \ V \circ g)(\x - \y)$, where $g(\x)$ is the smearing function in Eq. \eqref{gaussiansmearing}.

To guarantee that Eq.~\eqref{massdI} represents a good choice for $\hat \mu_\text{I}(\x)$, we make a comparison with the dissipative KTM model in the appropriate limit. In particular, we substitute Eq.~\eqref{massdR} and Eq.~\eqref{massdI} in Eq.~\eqref{TDmasterdissip1}, and we rewrite the position operator as in Eq.~\eqref{fluttuazioni}. In the assumption of small quantum fluctuations, we can take the linear limit of the dissipative TD master equation~\eqref{TDmasterdissip1}, which reads
\bqali
\label{linearmasterTD}
\frac{\D}{\D t}\hat\rho_{t}  &= - \frac{i}{\hbar}[\hat{H}_{0} + \hat{H}_\text{grav},\hat\rho_{t}] \\
& - \sum_{k,j=1}^{N}\sum_{l,n=1}^{3}\frac{Gm_{k}m_{j}\eta_{kjln}}{2\hbar}[\hat{x}_{kl},[\hat{x}_{jn},\hat\rho_{t}]] \\
& -\sum_{k,j=1}^{N}\sum_{l,n=1}^{3}\frac{Gm_{k}m_{j}\alpha_{k}\alpha_{j}\eta_{kjln}}{4\hbar^{3}}[\hat{p}_{kl},[\hat{p}_{jn},\hat\rho_{t}]] \\
& - \sum_{k,j=1}^{N}\sum_{l,n=1}^{3}\frac{iGm_{k}m_{j}\alpha_{j}\eta_{kjln}}{2\hbar^{2}}[\hat{x}_{kl},\{\hat{p}_{jn},\hat\rho_{t}\}] \\
& - \sum_{k,j=1}^{N}\sum_{l,n=1}^{3}\frac{Gm_{k}m_{j}\alpha_{j}\eta_{kjln}}{2\hbar^{2}}[\hat{x}_{kl},[\hat{p}_{jn},\hat\rho_{t}]],
\eqali
where $\hat{x}_{kl}$ and $\hat{p}_{kl}$ are the components in the $l$-th direction of $\Delta\hat{\x}_{k}$ and $\Delta\hat{\p}_{k}$ respectively, and
\begin{equation}
\hat{H}_\text{grav} = \frac{G}{4}\sum_{k,j=1}^{N}\sum_{l,n=1}^{3}m_{k}m_{j}\eta_{kjln}(\hat x_{kl}-\hat{x}_{jl})(\hat x_{kn}-\hat{x}_{jn}),
\end{equation}
is the gravitational interaction in the linear limit. Here, we defined
\bqali\label{def.eta}
\eta_{kjln} = \int \frac{\D^{3}q}{2\pi^{2}\hbar^{3}}\,\frac{\tilde{g}^{2}(\qq)}{q^2} q_{l}q_{n}\, e^{\frac{i}{\hbar}\qq\cdot(\x_{k}^{(0)} - \x_{j}^{(0)})}, 
\eqali
where we made explicit the Fourier transform of the Newtonian gravitational potential $\tilde{V}(\qq) = -{4\pi G\hbar^{2}}/{q^{2}}$. 

Now, we reduce the problem to that of only two harmonic oscillators at frequency $\omega$ in one dimension, with $m_{k}  = m$ and $\alpha_{k} = \alpha$. Thus, 
Eq.~\eqref{linearmasterTD} becomes
\bqali
\label{linearmasterTD2}
&\frac{\D}{\D t}\hat\rho_{t}  = - \frac{i}{\hbar}[\hat{H},\hat\rho_{t}] - \sum_{k,j=1}^{2}\frac{iGm^{2}\alpha\eta_{kj}}{2\hbar^{2}}[\hat{x}_{k},\{\hat{p}_{j},\hat\rho_{t}\}]\\
& - \!\sum_{k,j=1}^{2}\!\frac{Gm^{2}\eta_{kj}}{2\hbar}[\hat{x}_{k},[\hat{x}_{j},\hat\rho_{t}]] -\! \sum_{k,j=1}^{2}\!\frac{Gm^{2}\alpha\eta_{kj}}{2\hbar^{2}}[\hat{x}_{k},[\hat{p}_{j},\hat\rho_{t}]]\\&-\sum_{k,j=1}^{2}\frac{Gm^{2}\alpha^{2}\eta_{kj}}{4\hbar^{3}}[\hat{p}_{k},[\hat{p}_{j},\hat\rho_{t}]],  \\
\eqali
where $\eta_{kj}=\eta_{kj11}$ are explicitly computed in Appendix \ref{appendixE}, and
\begin{equation}
\label{TDhamilton}
\hat{H} = \sum_{k=1}^{2}\left(\frac{\hat{p}_{k}^{2}}{2m} + \frac{m\Omega^{2}}{2}\hat{x}_{k}^{2}\right) - Gm^{2}\eta_{12}\hat{x}_{1}\hat{x}_{2},
\end{equation}
includes also the linearized quantum gravitational interaction $\hat H_\text{grav}$ and we defined
\begin{equation}
\label{Omega}
\Omega^{2} = \omega^{2} + Gm\eta_{12}.
\end{equation}
We notice that Eq.~\eqref{linearmasterTD2} has a structure similar to that of the dissipative KTM master equation \eqref{dissipationrho}, 
although there are some some differences. In particular, additional terms
in the commutators and anticommutators mixing the position and momentum operators of different particles appear, and the coefficients $\eta_{kj}$ differ. 

By inserting the explicit expressions of $\eta_{kj}$ [cf.~Eq.~\eqref{eta112m}] in Eq.~\eqref{linearmasterTD2}, we can decouple the center-of-mass and relative dynamics. Correspondingly, Eq.~\eqref{linearmasterTD2} can be divided in two independent master equations. That for the center-of-mass reads
\bqali
\label{centerofmass}
&\frac{\D}{\D t}\hat{\rho}_\text{cm}  = -\frac{i}{\hbar}[\hat{H}_\text{cm},\hat{\rho}_\text{cm}] - \frac{Gm^{2}}{\hbar}\eta_{+}[\hat{x}_\text{cm},[\hat{x}_\text{cm},\hat{\rho}_\text{cm}]] \\
& - \frac{Gm^{2}\alpha^{2}}{8\hbar^{3}}\eta_{+}[\hat{p}_\text{cm},[\hat{p}_\text{cm},\hat{\rho}_\text{cm}]] \\
&- \frac{iGm^{2}\alpha}{2\hbar^{2}}\eta_{+} [\hat{x}_\text{cm},\{\hat{p}_\text{cm},\hat{\rho}_\text{cm}\}] \\
& - \frac{Gm^{2}\alpha }{2\hbar^{2}}\eta_{+} [\hat{x}_\text{cm},[\hat{p}_\text{cm},\hat{\rho}_\text{cm}]],
\eqali
where $\eta_{+} = \eta + \eta_{12}$ with $\eta = (6\sqrt{\pi}R_{0}^{3})^{-1}$ and we defined 
\begin{equation}
\label{harmoscillcm}
\hat{H}_\text{cm} = \frac{\hat{p}_\text{cm}^{2}}{4m} + m\Omega_\text{cm}^{2}\hat{x}_\text{cm}^{2},
\end{equation}
with $\Omega_\text{cm}^{2} = \Omega^{2} - Gm\eta_{12}=\omega^2$.
On the other hand, the relative dynamics is described by the following master equation
\bqali
\label{relative}
&\frac{\D}{\D t}\hat{\rho}_\text{rel}  = -\frac{i}{\hbar}[\hat{H}_\text{rel},\hat{\rho}_\text{rel}] - \frac{Gm^{2}}{4\hbar}\eta_{-}[\hat{x}_\text{rel},[\hat{x}_\text{rel},\hat{\rho}_\text{rel}]] \\
& - \frac{Gm^{2}\alpha^{2}}{2\hbar^{3}}\eta_{-}[\hat{p}_\text{rel},[\hat{p}_\text{rel},\hat{\rho}_\text{rel}]] \\
&- \frac{iGm^{2}\alpha}{2\hbar^{2}}\eta_{-} [\hat{x}_\text{rel},\{\hat{p}_\text{rel},\hat{\rho}_\text{rel}\}] \\
& - \frac{Gm^{2}\alpha}{2\hbar^{2}}\eta_{-} [\hat{x}_\text{rel},[\hat{p}_\text{rel},\hat{\rho}_\text{rel}]],
\eqali
where $\eta_{-} = \eta - \eta_{12}$ and
\bq
\label{harmoscillrel}
\hat{H}_{\text{rel}} = \frac{\hat{p}_{\text{rel}}^{2}}{m} + \frac{m}{4}\Omega_{\text{rel}}^{2}\hat{x}_{\text{rel}}^{2},
\eq
is the relative Hamiltonian with $\Omega_{\text{rel}}^{2} = \Omega^{2} + Gm\eta_{12}=\omega^{2} + 2Gm\eta_{12}$. The coefficients $\eta_{\pm}$ explicitly read
\bq
\eta_{\pm}=\frac{1}{2\sqrt{\pi}R_{0}^{3}}\left[\frac13\pm\frac{e^{-\frac{d^{2}}{4R_{0}^{2}}}}{d^{2}}(4R_{0}^{2} + d^{2})\right] \mp\frac{2{\erf}\left(\frac{d}{2R_{0}}\right)}{d^{3}}.
\eq
Then, by moving along the lines drawn in Sec.~\ref{tre}, we can analyze the behaviour of the asymptotic energy described by Eq.~\eqref{linearmasterTD2}. After lengthly calculations, which are reported in Appendix~\ref{appendixF},  
we obtain the asymptotic energy of the system, which reads
\bqali
\label{asyntHcmf}
\braket{\hat{H}}_{\infty} = \frac{\hbar^{2}}{m\alpha} + \frac{\alpha m\omega^{2}}{2}- \frac{\alpha m^{2}G}{2}\eta_{-} +\frac{G^{2}\alpha^{3}m^{5}}{4\hbar^{2}}(\eta^{2} + \eta_{12}^{2}).
\eqali
Since  $R_0$ is the spread of the smearing function regulating the Newtonian potential, we can assume that $R_0$ is small with respect to the distance $d$. Thus, in the limit of $R_{0}$ being small, one finds that $\eta_{12}\to -2/d^3$ and $\eta_{\pm}\to\mp2/d^3$. Correspondingly, one has $\eta^2+\eta_{12}^2\to4/d^6$.
By substituting these values, Eq.~\eqref{TDhamilton} becomes equal to the Hamiltonian of the KTM model, and Eq.~\eqref{asyntHcmf} equates the asymptotic energy of the dissipative KTM model in Eq.~\eqref{energiaasint}.  Finally, applying the  limit $\alpha\to 0$, one finds that  the system thermalizes at the temperature expressed in Eq.~\eqref{tefft}, namely $T_{\text{eff}} = {\hbar^{2}}/{2m_{0}\alpha_{0}\kB}$, where we employed Eq.~\eqref{alfa}.

\section{Conclusions}

The protocol based on the continuous measurement and feedback mechanism in the KTM and TD model well reconstructs the quantum gravitational interaction from a fundamentally classical description \cite{reyes2020gravitational}. The corresponding appearance of decoherence terms which lead to an indefinite energy increase is, however, a problem. Here, we suggested a way to account for this feature by suitably modifying the protocol. We derived the dissipative generalizations of the two models, and showed that -- in the appropriate limits -- the system under such a protocol thermalizes to an effective temperature [cf.~Eq.~\eqref{tefft}].

With our generalization, the energy of the system remains finite also asymptotically. Yet, energy conservation at each time is still lacking. A possible step forward in such a direction could be to \textit{upgrade} the stochastic noises of the protocol to physical dynamical fields \cite{adler2004quantum, pearle2000wavefunction}. Then, one could in principle be able to conserve the total energy of the system plus the stochastic field. This is subject for future research.

\acknowledgments

MC is supported by UK EPSRC (grant nr.~EP/T028106/1).
AB acknowledges financial support from the INFN, the University of Trieste and the support by grant number (FQXi-RFP-CPW-2002) from the Foundational Questions Institute and Fetzer Franklin Fund, a donor advised fund of Silicon Valley Community Foundation. MC and  AB acknowledge support from the H2020 FET Project TEQ (Grant No. 766900).

\appendix

\section{Continuous quantum measurement and feedback framework}\label{appendixA}

By following the approach in \cite{jacobs2006straightforward, wiseman2009quantum}, we briefly review the dynamics due to a continuous quantum measurement and due to the feedback. For the sake of simplicity we consider only the discrete case of a one-dimensional system made of two particles and then the case of a generic continuous system in three dimensions.

\subsection{One dimensional two-particle system}
\emph{i) Continuous measurement}.-- The stochastic Schr\"odinger equation of two particles due to a continuous quantum measurement of arbitrary operators $\hat{A}_{k}$ is given by
\bqali
\label{measure}
\D\ket{\psi_{t}}_{\text{m}} &  = -\sum_{k=1}^{2}\frac{\gamma_{k}}{8\hbar^{2}}\left(\hat{A}_{k}^{\dag}\hat{A}_{k} + \braket{\hat{A}_{k}^{\dagger}}_{t}\left(\braket{\hat{A}_{k}}_{t} - 2\hat{A}_{k}\right)\right)\ket{\psi_{t}}\D t \\
& +\sum_{k=1}^{2}\frac{\sqrt{\gamma_{k}}}{2\hbar}\left(\hat{A}_{k} -\braket{\hat{A}_{k}}_{t}\right)\ket{\psi_{t}}\D W_{k,t},
\eqali
where $\braket{\hat{A}_{k}}_{t} = \bra{\psi_{t}}\hat{A}_{k}\ket{\psi_{t}}$. The constants $\gamma_{k}$ represent the information rate of the measurement and $\D W_{k,t}$ are standard independent Wiener process such that $\mathbb{E}[\,\D W_{k,t}] = 0$ and $\mathbb{E}[\,\D W_{k,t}\D W_{j,t}] = \delta_{kj}\D t$.
The measurement record corresponding to the measurement of $\hat{A}_{k}$ is defined as
in Eq.~\eqref{m.record}.
This is a stochastic quantity centered at the average value $\frac{1}{2}\braket{\hat{A}_{k} + \hat{A}_{k}^{\dagger}}_{t}$ and with a variance defined by $\gamma_{k}$ and  $\D W_{k,t}$.

\emph{ii) Feedback dynamics}.-- After performing the measurement of the operators $\hat{A}_{k}$, we can send the corresponding measurement result to the complementary subsystem. This operation can be performed by employing the following feedback Hamiltonian
\begin{equation}
\label{feedback}
\hat{H}_{\text{fb}} = \chi_{1}r_{1}\hat{B}_{2} + \chi_{2}r_{2}\hat{B}_{1},
\end{equation}
where $\chi_{k}$ are real constants and $\hat B_k$ are suitable operators.

The corresponding feedback equation is computed by unitarily evolve the state $\ket{\psi_{t}}$ with respect to $\hat{H}_{\text{fb}} $. Then, the infinitesimal increment of the state is given by
\bqali
\D\ket{\psi_{t}}_{\text{fb}} & = -\sum_{{\substack{k,j=1\\j\ne k}}}^{2} \left[\left( \frac{i\chi_{k}}{2\hbar} \braket{\hat{A}_{k} + \hat{A}_{k}^{\dagger}}_{t} \hat{B}_{j} + \frac{\chi_{k}^{2}}{2\gamma_{k}} \hat{B}_{j}^{2} \right) \D t \right. \\
&\left. -\frac{i\chi_{k}}{\sqrt{\gamma_{k}}} \hat{B}_{j}\D W_{k,t}\right]\ket{\psi_{t}}.
\eqali

\emph{iii) Combined evolution}.--
Finally, the stochastic Schr\"odinger equation comprising the combined effects of the continuous measurement and feedback is given by
\begin{equation}\label{dpsi.combined}
\D \ket{\psi_{t}} = \D\ket{\psi_{t}}_{\text{m}} + \D\ket{\psi_{t}}_{\text{fb}} + \D\ket{\psi_{t}}_{\text{It\^{o}}},
\end{equation}
where the third term comes from the stochastic It\^{o} calculus:
\begin{equation}
\D\ket{\psi_{t}}_{\text{It\^{o}}} = -\sum_{{\substack{k,j=1\\j\ne k}}}^{2} \frac{i}{2\hbar}\chi_{k} \hat{B}_{j} (\hat{A}_{k} - \braket{\hat{A}_{k}}_{t})\ket{\psi_{t}}\D t.
\end{equation}
By fixing $\hat A_k=\hat x_k$ and $\hat B_j=\hat x_j$, the latter expression gives Eq.~\eqref{continuousmeasure}.

The master equation for the density matrix corresponding to Eq.~\eqref{dpsi.combined} is found by performing the stochastic average over the noise, and is given by
\bqali
\label{mastereq}
&\frac{\D}{\D t}\hat\rho_{t}  = -\frac{i}{\hbar}[\hat{H}_{0},\hat\rho_{t}] - \sum_{{\substack{k,j=1\\j\ne k}}}^{2}\frac{\chi_{k}^{2}}{2\gamma_{k}}[\hat{B}_{j},[\hat{B}_{j},\hat\rho_{t}]]\\
&- \sum_{{\substack{k,j=1\\j\ne k}}}^{2} \frac{i}{2\hbar}\chi_{k}[\hat{B}_{j},\hat{A}_{k}\hat\rho_{t} + \hat\rho_{t}\hat{A}_{k}^{\dagger}]  \\
& +\sum_{k=1}^{2}\frac{\gamma_{k}}{4\hbar^{2}}\left[\hat{A}_{k}\hat\rho_{t}\hat{A}_{k}^{\dagger} - \frac{1}{2}\{\hat{A}_{k}^{\dagger}\hat{A}_{k},\hat\rho_{t}\}\right],
\eqali
where we added the unitary evolution described by the Hamiltonian $\hat{H}_{0}$. The second and third terms of Eq.~\eqref{mastereq} come from the feedback dynamics, while the last one comes from continuous measurement of $\hat{A}_{k}$. Moreover, by suitably rearranging the third term as
\bq
 [\hat{B}_{j}\hat{A}_{k} + \hat{A}_{k}^{\dagger}\hat{B}_{j},\hat\rho_{t}] + [\hat{B}_{j}\hat\rho_{t},\hat{A}_{k}^{\dagger} ] + [\hat\rho_{t}\hat{B}_{j},\hat{A}_{k}] ,
\eq
one can single out an effective correction to the Hamiltonian. Such a correction reads
\begin{equation}
\hat{H}_{\text{I}} = \sum_{{\substack{k,j=1\\j\ne k}}}^{2} \frac{\chi_{k} }{2}\left(\hat{B}_{j}\hat{A}_{k} + \hat{A}_{k}^{\dagger}\hat{B}_{j}\right),
\end{equation}
and it can be interpreted as a quantum interaction Hamiltonian between the two particles. We underline that such a term comes from the combined action of the continuous measurement and feedback dynamics.

\subsection{Continuous system in three dimensions}

\emph{i) Continuous measurement}.-- The stochastic Schr\"odinger equation resulting from the continuous measurement of an arbitrary operator $\hat{A}(\x)$ is
\bqali
&\D\ket{\psi_{t}}_{\text{m}}  = -\int \frac{\D^{3}x\D^{3}y}{8\hbar^{2}}  \gamma(\x - \y) \left[\hat{A}^{\dagger}(\x)\hat{A}(\y) \right.\\
&\left. + \braket{\hat{A}^{\dagger}(\x)}_{t}\left(\braket{\hat{A}(\y)}_{t} - 2\hat{A}(\y)\right)\right]\ket{\psi_{t}}\D t \\
& + \int \frac{\D^{3}x }{2\hbar} \left(\hat{A}(\x) - \braket{\hat{A}(\x)}_{t}\right)\ket{\psi_{t}}\D W_{t}(\x),
\eqali
where the noise is now described by $\mathbb{E}[\,\D W_{t}(\x)] = 0$ and $\mathbb{E}[\,\D W_{t}(\x)\D W_{t}(\y)] = \gamma(\x - \y)\D t$, where $\gamma(\x - \y)$ is an arbitrary spatial correlator. The measurement record reads as in Eq.~\eqref{recordsistemacontinuo},
where $\gamma^{-1}(\x - \y)$ is the inverse function of $\gamma(\x - \y)$. The following expression
\begin{equation}
(\gamma \circ \gamma^{-1})(\x - \y) = \int \D^{3}r\,\gamma(\x - {\bf r})\gamma^{-1}({\bf r} - \y) = \delta(\x - \y),
\end{equation}
relates these two functions.

\emph{ii) Feedback dynamics}.-- The measurement record is broadcasted to the system by using the following feedback Hamiltonian
\begin{equation}
\hat{H}_{\text{fb}} = \int \D^{3}x\D^{3}y\, V(\x - \y) \hat{B}(\x)r(\y),
\end{equation}
where $\hat{B}(\x)$ is an Hermitian operator and $V(\x - \y)$ is a generic function. The corresponding infinitesimal state evolution reads
\bqali
\D\ket{\psi_{t}}_{\text{fb}} & = -\int \D^{3}x\D^{3}y\,\left[\frac{i}{2\hbar}V(\x - \y)\braket{\hat{A}(\x) + \hat{A}^{\dagger}(\x)}_{t} \D t \right.\\
& + \frac{1}{2}(V\circ \gamma^{-1}\circ V)(\x - \y)\hat{B}(\y)\D t \\
&\left. + i(V\circ \gamma^{-1})(\x - \y)\D W_{t}(\y)\right]\hat{B}(\x)\ket{\psi_{t}}.
\eqali

\emph{iii) Combined evolution}.--
The stochastic Schr\"odinger equation of the combined dynamics is given by Eq.~\eqref{dpsi.combined}, where
\bqali
\D\ket{\psi_{t}}_{\text{It\^{o}}} = - i\int \frac{\D^{3}x\D^{3}y}{2\hbar}V(\x - \y)\hat{B}(\x)\\
\left(\hat{A}(\y) - \braket{\hat{A}(\y)}_{t}\right)\ket{\psi_{t}}\D t.
\eqali
Correspondingly, the master equation reads
\bqali
\label{generalmastercontinua}
& \frac{\D}{\D t}\hat\rho_{t} = -\frac{i}{\hbar}[\hat{H}_{0},\hat\rho_{t}] \\ 
& -i\int \frac{\D^{3}x\D^{3}y}{2\hbar} V(\x - \y)[\hat{B}(\x),\hat{A}(\y)\hat\rho_{t}+ \hat\rho_{t}\hat{A}^{\dagger}(\y)] \\
& - \int \frac{\D^{3}x\D^{3}y}{2} (V\circ \gamma^{-1} \circ V)(\x - \y)[\hat{B}(\x), [\hat{B}(\y),\hat\rho_{t}]]\\
& + \int \frac{\D^{3}x\D^{3}y}{4\hbar^{2}} \gamma(\x - \y)\left[\hat{A}(\x)\hat\rho_{t}\hat{A}^{\dagger}(\y) - \frac{1}{2}\{\hat{A}^{\dagger}(\x)\hat{A}(\y),\hat\rho_{t}\}\right] 
\eqali
We notice that Eq.~\eqref{generalmastercontinua} has the same form of Eq.~\eqref{mastereq}. The second and third terms come from the feedback dynamics, while the last term comes from continuous measurement. Similarly as done previously, we can extract an interaction Hamiltonian reading
\begin{equation}\label{eq.feedbak.cont}
\hat{H}_\text{I} =\frac12 \int {\D^{3}x\D^{3}y} V(\x - \y)\left(\hat{B}(\x)\hat{A}(\y) + \hat{A}^{\dagger}(\y)\hat{B}(\x)\right),
\end{equation}
by suitably rearrange the second term of the master equation.
Thus, also in the case of a continuous system, one obtains an interaction Hamiltonian term from the combined action of continuous measurement and feedback dynamics.

\section{Calculation of the $\eta_{kj}$ coefficients in the linearized dissipative TD master equation}\label{appendixE}

Here, we compute the coefficients $\eta_{kj}$ appearing in Eq.~\eqref{linearmasterTD2}. In particular, they are derived from Eq.~\eqref{def.eta} by restricting the problem to the one dimensional case. Correspondingly, we find
\bqali
\label{etam}
\eta_{kj} = \int \frac{\D^{3}q}{2\pi^{2}\hbar^{3}} \frac{\tilde{g}^{2}(\qq)}{q^2}q_{1}^{2}\ e^{\frac{i}{\hbar}q_{1}(x_{k}^{(0)} - x_{j}^{(0)})}.
\eqali
Now, we can compare Eq.~\eqref{linearmasterTD2} with the KTM master equation \eqref{dissipationrho} once computing the coefficients in Eq.~\eqref{etam}, which can be done by moving to spherical coordinates and choosing a suitable form of the smearing function. In particular, when considering the Gaussian smearing in Eq. \eqref{gaussiansmearing}, whose Fourier transform reads $\tilde{g}(\qq) = e^{-\frac{q^{2}R_{0}^{2}}{2\hbar^{2}}}$, we find that the coefficients become
\bqali
\label{eta112m}
&\eta_{kk} = \eta,\quad \text{for}\quad k = 1,2,\\
&\eta_{12} =\eta_{21}=\frac{1}{2\sqrt{\pi}R_{0}^{3}}\frac{e^{-\frac{d^{2}}{4R_{0}^{2}}}}{d^{2}}(4R_{0}^{2} + d^{2}) -\frac{2{\erf}\left(\frac{d}{2R_{0}}\right)}{d^{3}},
\eqali
where $\eta = (6\sqrt{\pi}R_{0}^{3})^{-1}$ and $d=|x_1^{(0)}-x_2^{(0)}|$. 

\section{Calculation of the mean energy in the dissipative TD model}\label{appendixF}

By following the approach used for the dissipative KTM model, we compute the evolution of
$\hat{T}_\text{cm} = {\hat{p}_\text{cm}^{2}}/{4m}$ and $\hat{V}_\text{cm} = m\Omega_\text{cm}^{2}\hat{x}_\text{cm}^{2}$ using Eq. \eqref{centerofmass}. We find the following system of differential equations
\bqali
\label{centermassdiff2}
\frac{\D}{\D t}\braket{\hat{V}_\text{cm}}_{t} & = \frac{\Omega_\text{cm}^{2}}{2}\braket{\{\hat{p}_\text{cm},\hat{x}_\text{cm}\}}_{t} + \frac{Gm^{3}\alpha^{2}\Omega_\text{cm}^{2}\eta_{+}}{4\hbar},  \\
\frac{\D}{\D t}\braket{\hat{T}_\text{cm}}_{t} & = - \frac{2Gm^{2}\alpha\eta_{+}}{\hbar}\braket{\hat{T}_\text{cm}}_{t}  \\
& - \frac{\Omega_\text{cm}^{2}}{2}\braket{\{\hat{p}_\text{cm},\hat{x}_\text{cm}\}}_{t} + \frac{\hbar Gm\eta_{+}}{2}, \\
\frac{\D}{\D t}\braket{\{\hat{p}_\text{cm},\hat{x}_\text{cm}\}}_{t} & = - \frac{Gm^{2}\alpha\eta_{+}}{\hbar}\braket{\{\hat{p}_\text{cm},\hat{x}_\text{cm}\}}_{t} + 4\braket{\hat{T}_\text{cm}}_{t}   \\
&- 4\braket{\hat{V}_\text{cm}}_{t} -Gm^{2} \alpha\eta_{+}.
\eqali
Similarly, we compute the evolution of
$\hat{T}_\text{rel} = {\hat{p}_\text{rel}^{2}}/{m}$ and $\hat{V}_\text{rel} = \frac{m}{4}\Omega_\text{rel}^{2}\hat{x}_\text{rel}^{2}$ using Eq.~\eqref{relative}. The energy behaviour is described by
\bqali
\label{h12time}
\frac{\D}{\D t}\braket{\hat{V}_\text{rel}}_{t} & = \frac{\Omega_\text{rel}^{2}}{2}\braket{\{\hat{p}_\text{rel},\hat{x}_\text{rel}\}}_{t} + \frac{Gm^{3}\alpha^{2}\Omega_\text{rel}^{2}\eta_{-}}{4\hbar},  \\
\frac{\D}{\D t}\braket{\hat{T}_\text{rel}}_{t} & = - \frac{2Gm^{2}\alpha\eta_{-}}{\hbar}\braket{\hat{T}_\text{rel}}_{t}  \\
& - \frac{\Omega_\text{rel}^{2}}{2}\braket{\{\hat{p}_\text{rel},\hat{x}_\text{rel}\}}_{t} + \frac{\hbar Gm\eta_{-}}{2}, \\
\frac{\D}{\D t}\braket{\{\hat{p}_\text{rel},\hat{x}_\text{rel}\}}_{t} & = - \frac{Gm^{2}\alpha\eta_{-}}{\hbar}\braket{\{\hat{p}_\text{rel},\hat{x}_\text{rel}\}}_{t} + 4\braket{\hat{T}_\text{rel}}_{t}   \\
&- 4\braket{\hat{V}_\text{rel}}_{t} -Gm^{2} \alpha\eta_{-}.
\eqali
By imposing that the time derivatives in Eq.~\eqref{centermassdiff2} and Eq.~\eqref{h12time} vanish, we find
\bqali
\braket{\{\hat{p}_\text{cm},\hat{x}_\text{cm}\}}_{\infty} & = -\frac{Gm^{3}\alpha^{2}\eta_{+}}{2\hbar} \\
\braket{\{\hat{p}_\text{rel},\hat{x}_\text{rel}\}}_{\infty} &= -\frac{Gm^{3}\alpha^{2}\eta_{-}}{2\hbar} \\ 
\braket{\hat{T}_\text{cm}}_{\infty} & = \frac{\hbar^{2}}{4m\alpha} + \frac{\alpha m\Omega_\text{cm}^{2}}{8},\\
\braket{\hat{T}_\text{rel}}_{\infty} & = \frac{\hbar^{2}}{4m\alpha} + \frac{\alpha m\Omega_\text{rel}^{2}}{8}, \\
\braket{\hat{V}_\text{cm}}_{\infty} & = \braket{\hat{T}_\text{cm}}_{\infty} + \frac{G^{2}m^{5}\alpha^{3}\eta_{+}^{2}}{8\hbar^{2}} - \frac{Gm^{2}\alpha\eta_{+}}{4}, \\
\braket{\hat{V}_\text{rel}}_{\infty} & = \braket{\hat{T}_\text{rel}}_{\infty} + \frac{G^{2}m^{5}\alpha^{3}\eta_{-}^{2}}{8\hbar^{2}} - \frac{Gm^{2}\alpha\eta_{-}}{4},
\eqali
which gives Eq.~\eqref{asyntHcmf} once considering that
$\hat{H} = \hat{H}_\text{cm} + \hat{H}_\text{rel} = \hat{T}_\text{cm} + \hat{V}_\text{cm} + \hat{T}_\text{rel} + \hat{V}_\text{rel}$, and $\Omega_\text{cm}^{2} = \Omega^{2} - Gm\eta = \omega^{2}$, $\Omega_{\text{rel}}^{2} = \Omega^{2} + Gm\eta_{12}=\omega^{2} + 2Gm\eta_{12}$, $\eta_{\pm} = \eta \pm \eta_{12}$.


%

\end{document}